\documentclass[a4paper]{article}

\usepackage{icrc2013}

\title{The Site of the ASTRI SST-2M Telescope Prototype}

\shorttitle{ASTRI SST-2M Prototype: The Site}

\authors{
Maria Concetta Maccarone$^{1}$,
Giuseppe Leto$^{2}$,
Pietro Bruno$^{2}$,
Mauro Fiorini$^{3}$,
Alessandro Grillo$^{2}$,
Alberto Segreto$^{1}$,
Luca Stringhetti$^{3}$,
for the ASTRI Collaboration$^{4}$.
}

\afiliations{
$^1$ INAF - IASF Palermo, Via U. La Malfa 153, I-90146 Palermo, Italy \\
$^2$ INAF - Osservatorio Astrofisico di Catania, Via S. Sofia 78, I-95123 Catania, Italy \\
$^3$ INAF - IASF Milano, Via E. Bassini 15, I-20133 Milano, Italy \\
$^4$ http://www.brera.inaf.it/astri\\
}

\email{Cettina.Maccarone@iasf-palermo.inaf.it}

\abstract{ASTRI is a Flagship Project financed by the Italian Ministry of Education, University and Research, and led by the Italian National Institute of Astrophysics, INAF. Primary goal of the ASTRI project is the design and production of an end-to-end prototype of Small Size Telescope for the CTA (Cherenkov Telescope Array) in a dual-mirror configuration (SST-2M) equipped with a camera at the focal plane composed by an array of Silicon Photo-Multipliers and devoted to the investigation of the highest gamma-ray energy band. The ASTRI SST-2M prototype will be placed at the INAF "M.G. Fracastoro" observing station in Serra La Nave on the Etna Mountain near Catania, Italy. After the verification tests, devoted to probe the technological solutions adopted, the ASTRI SST-2M prototype will perform scientific observations on the Crab Nebula and on some of the brightest TeV sources. Here we present the Serra La Nave site, its meteorological and weather conditions, the sky darkness and visibility, and the complex of auxiliary instrumentation that will be used on site to support the calibration and science verification phase as well as the regular data reconstruction and analysis of the ASTRI SST-2M prototype.}


\begin{document}
\maketitle

%Begin a section.
\section{Introduction}

The next generation of the Imaging Atmospheric Cherenkov Telescopes will explore the uppermost end of the Very High Energy domain with unprecedented sensitivity, angular resolution and imaging quality: this is the ambitious target of the Cherenkov Telescope Array, CTA \cite{bib:CTA,bib:CTA_2013} which plans the construction of about one hundred of telescopes divided in three kinds of configurations in order to cover the energy range from tens of GeV up to 100 TeV and beyond.
The Italian contribution in this field is represented by the ASTRI program, a "Flagship Project" financed by the Italian Ministry of Education, University and Research, and led by INAF, the Italian National Institute of Astrophysics.

Primary goal of the ASTRI (Astrofisica con Specchi a Tecnologia Replicante Italiana) program is the design and production of an end-to-end prototype of a Small Size Telescope \cite{bib:ASTRI,bib:Pareschi} for the CTA and devoted, with its wide full field of view of about 10 degrees, to the highest gamma-ray energy region. The telescope, named ASTRI SST-2M, is characterized by innovative technological solutions: for the first time in the design of Cherenkov telescopes it adopts together an optical system in dual-mirror configuration \cite{bib:Canestrari} and a camera at the focal plane composed by a matrix of multi-pixels Silicon Photo Multipliers \cite{bib:Catalano}.

The ASTRI SST-2M prototype will be tested on field:
the verification and scientific calibration phase will be performed in Italy. The telescope will be placed at the INAF "M.G. Fracastoro" observing station located in Serra La Nave on the Etna Mountain near Catania, Italy. The installation is foreseen in mid-2014, immediately followed by the start of the data acquisition. Here we present the Serra La Nave site, its weather and sky conditions, and the complex of auxiliary instrumentation that will be used on site to support the calibration and science verification phase as well as the regular data reconstruction and analysis of the ASTRI SST-2M prototype.

\section{The Serra La Nave Observing Station}

Apart from the obvious geophysical conditions needed to detect VHE gamma-rays using the Cherenkov observational approach, the site where install our ASTRI SST-2M telescope must satisfy several requirements as, for example, those concerning
atmospheric and meteorological conditions, accessibility and infrastructures. For this purpose,
we performed a detailed review of the INAF candidate \-ob\-ser\-ving stations present in the National territory \cite{bib:Maccarone_TN_Site}. The final choice was in favor of the Serra La Nave (SLN) site,
whose figures of merit are compliant with the general specification required for SST telescopes by the CTA Collaboration \cite{bib:CTA}, obviously restricted to the case of a single telescope. Moreover, the SLN site is located a short distance (about 30 km) from the laboratories of the Catania Astrophysical Observatory where the ASTRI SST-2M camera will be characterized and assembled.

 \begin{figure*}[!t]
  \centering
  \includegraphics[width=0.9\textwidth]{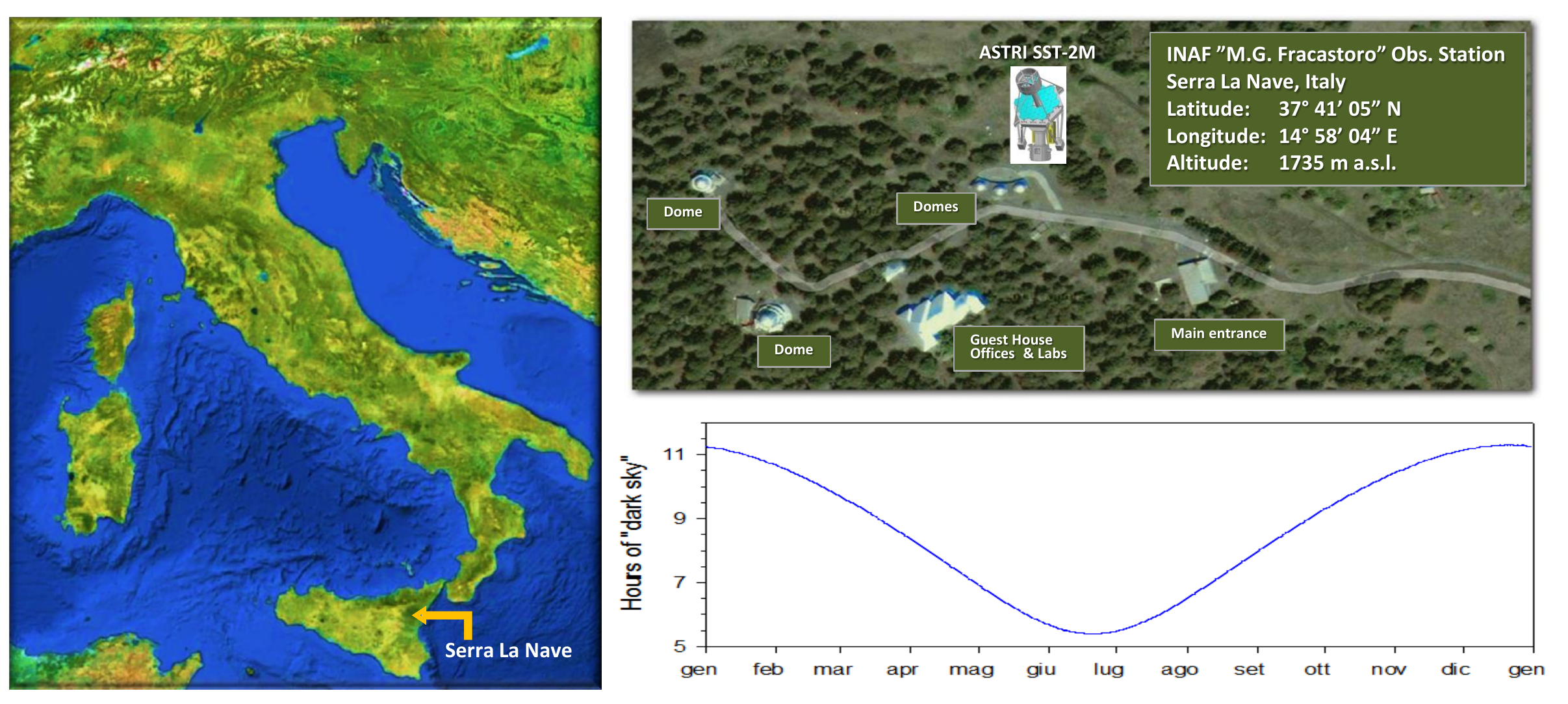}
  \caption{Location of Serra La Nave (left panel) and aerial view of the INAF "M.G. Fracastoro" observing station (upper right panel); superimposed is the ASTRI SST-2M sketch. The lower right panel shows the time extension of "dark nights" at Serra La Nave \cite{bib:UVscope_suite}.}
  \label{figure.1}
 \end{figure*}

The Serra La Nave observing station is at 1735 m a.s.l, 37$^\circ$ 41' 05" N Latitude, 14$^\circ$ 58' 04" E Longitude. The station is inside the "Parco dell'Etna" on the southern side of the Etna Volcano (Figure \ref{figure.1}), in a land protected from strong wind and from much of the fallout of volcanic ash which influences the transparency of the atmosphere. Although the region is defined with medium risk of seismicity, no earth-faults are present in territory so that the earthquake risk is strongly reduced \cite{bib:Leto_IR_seismic}. Tremors due to the Etna activity are modest, although several and frequent eruptive events have been registered during this year; volcanic ash can be present, depending on the wind direction, but only in minor content and for only few days per year. Ultimately, the SLN observing station has been identified as the best INAF Italian site for the installation, calibration and scientific verification of the ASTRI SST-2M Cherenkov telescope.

The main features of SLN, from the scientific observational point of view, can be summarized as follows:
\begin{itemize}
\item[$-$] horizon clear above 20$^\circ$ on average but sufficient, together with the high altitude and low latitude, to verify the performance of our Cherenkov prototype;
\item[$-$] medium level of light pollution extending in South-South-East direction under 30$^\circ$ of elevation;
\item[$-$] atmosphere transparency and relative humidity values well inside the limits required to perform Cherenkov observations;
\item[$-$] fraction of useful nights more than 53\%;
\item[$-$] time extension of "dark nights" \footnotemark  \footnotetext{dark night: time interval from the end of astronomical twilight in the evening to the beginning of astronomical twilight in the morning, when the Sun is under 18$^\circ$ below the horizon.} from about 6 hours (July) to more than 11 hours (January) \cite{bib:UVscope_suite}.
\end{itemize}

\begin{figure*}[!t]
  \centering
  \includegraphics[width=0.9\textwidth]{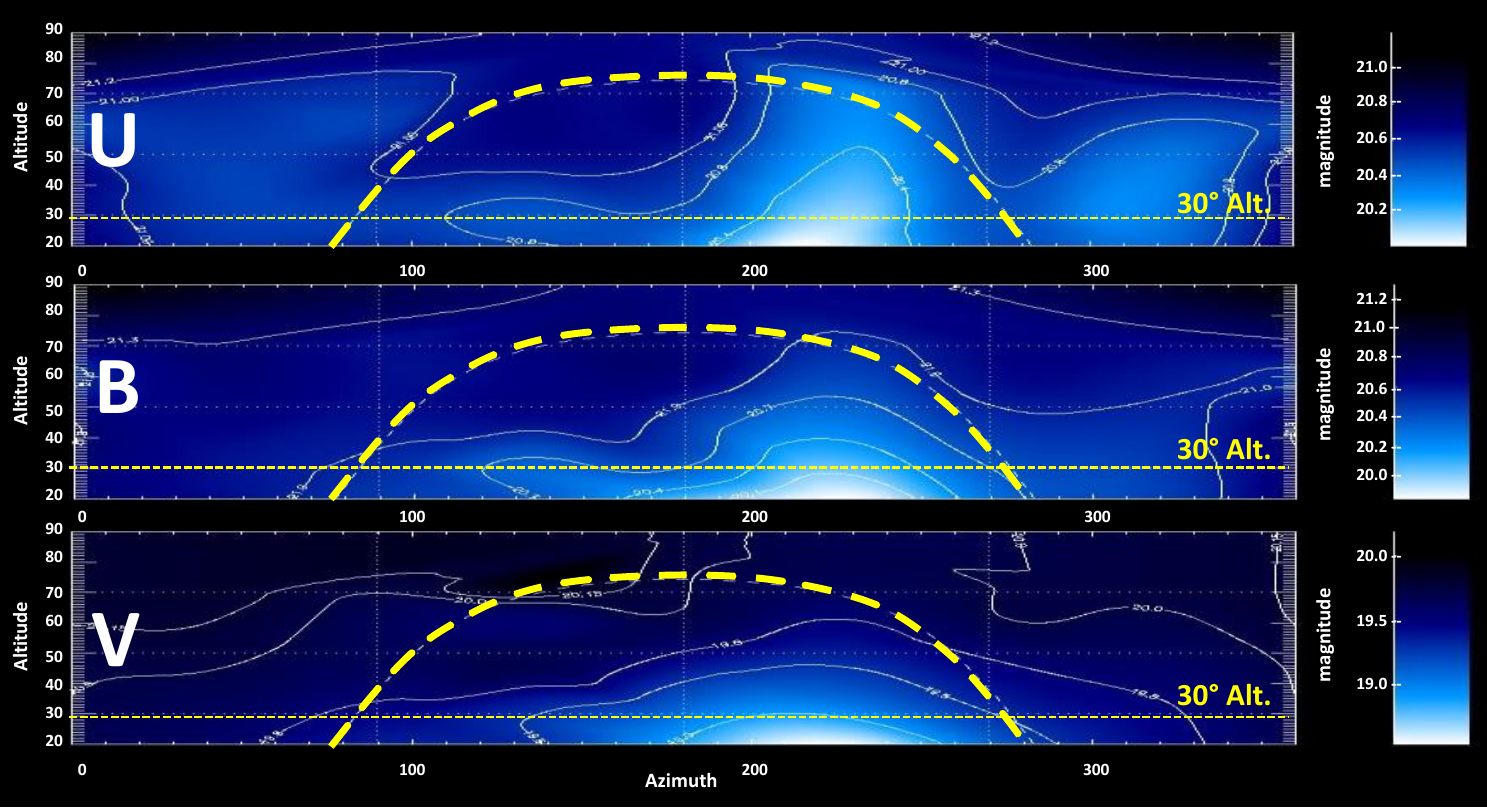}
  \caption{The sky brightness under U, B and V filters (see text for details).}
  \label{figure.2}
 \end{figure*}

Figure \ref{figure.2}  shows, with different color scales, the behavior of the sky brightness as registered with U, B and V filters. An evenly spaced  grid of alt-azimuth observations has been performed with the robotic APT2 telescope at SLN during a summer night (5 July 2011, three days after New Moon, 13$^\circ$C mean temperature, 63\% relative humidity). After a full reduction to the standard UBV Johnson system the sky brightness has been measured in mag/arcsec$^{2}$. It is evident the light pollution due to the residential areas in South-South-East direction under 30$^\circ$ elevation; nevertheless, the magnitude does not fall below 20.0, 19.8, 18.5 in U, B, and V, respectively. A source close to the Crab Nebula (marked as a dash line in the maps) in its apparent path as seen from the SLN site, is observed on a sky whose magnitude in U is less bright than 20.4; in the case of B, the magnitude will be greater than 20.7, while in the case of V is maintained near the 20.0 for a large part of the trajectory.

The analysis of the historical weather data in the period 2007 - 2013 April allows us to conclude that:

\begin{itemize}
\item[$-$] the temperature during the year ranges from -10$^\circ$C to 30$^\circ$C, max;
\item[$-$] the average relative humidity is 67\% in summer, and 79\% in winter;
\item[$-$] the wind speed is 7 km/h on average, with very-rare gusts (78 km/h max).
\end{itemize}

 \begin{figure}[t]
  \centering
  \includegraphics[width=0.49\textwidth]{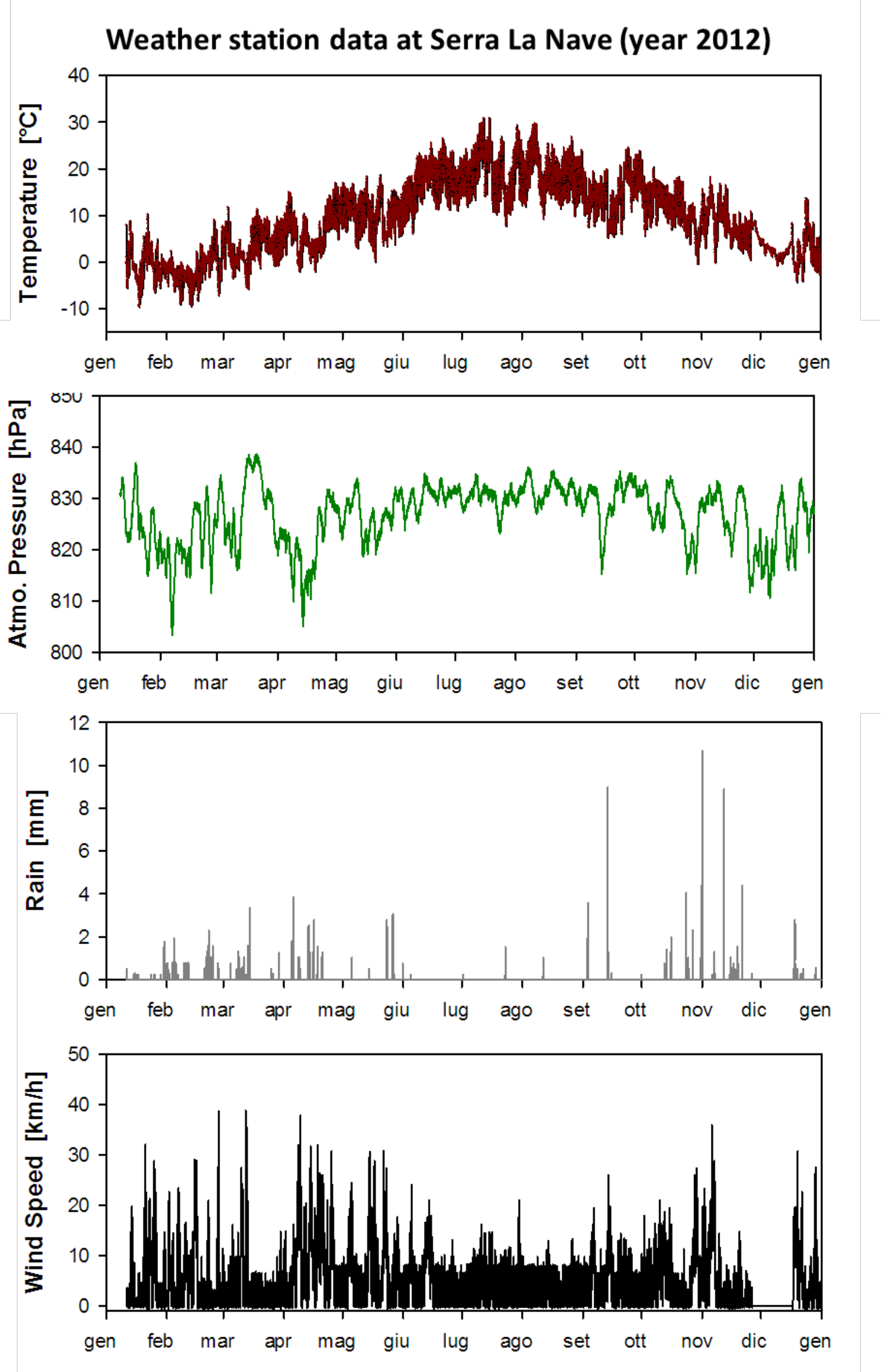}
  \caption{Profiles of temperature, pressure, rain and wind speed (averaged in 15 minutes time window) at Serra La Nave during the year 2012.}
  \label{figure.3}
 \end{figure}

The SLN logistic/functional aspects can be identified as follows:
\begin{itemize}
\item[$-$] the station is accessible during the entire year and has a permanent guardian. The access by road does not present problems in winter; during the last two years, the access by road was not available only for a maximum of four days per year due to the snow;
\item[$-$] the station is active since many years and pro\-vi\-ded with various instrumentation to be used before and during all the installation, calibration and regular data-taking phases of the ASTRI SST-2M prototype. Among them, a basic workshop for mechanics, electronics and optics purposes;
\item[$-$] manpower is available on site and personnel units can assist during the observations;
\item[$-$] the station is provided with housing for the technical and scientific staff dedicated to the various phases of installation, testing and operation of the ASTRI SST-2M prototype;
\item[$-$] the station is provided with public internet access, optical fiber and wireless everywhere;
\item[$-$] the station is located a short distance ($<2$ km) from inhabited areas and from safety/emergency structures ($<15$ km);
\item[$-$] the station is located a short distance ($\sim30$ km) from the laboratories of the Catania Astrophysical Observatory where the ASTRI SST-2M camera will be characterized and assembled.
\end{itemize}

\section{The Auxiliary Instrumentation}
The term "auxiliary" here refers to the instrumentation, installed at the Serra La Nave site and external to the ASTRI SST-2M prototype, which provides the necessary support for the monitoring of the operating conditions and for the calibration of the telescope; eventually, the auxiliary data will be archived in a proper database and will play their role crucial during the data reduction and analysis phases \cite{bib:Antonelli}. The main outcomes from the auxiliary equipment concern weather and environmental information, including sky brightness and atmosphere attenuation.

The general rule, compliant with the CTA requirements \cite{bib:CTA}, is that the ASTRI SST-2M prototype will perform observations unless the weather does not allow them; this control is managed by the continuously active weather station, a Vantage Pro2 Davis Instrument equipped with a Weather-Link Streaming Data Logger. Several parameters can be registered at the desired frequency starting from every 2 se\-conds; among them, temperature, humidity, pressure, wind speed and direction, rain and rain rate (some of them are shown in Figure \ref{figure.3} as acquired at SLN during year 2012). The WeatherLink will provide alarm output for all the necessary parameters, commanding the switch-off of all the telescope functions when any alert will be present.

The sky conditions in SLN will be primarily monitored by a fish-eye all-sky camera and a sky quality meter.

The all-sky camera chosen is the SBIG AllSky-340C color fish-eye model which provides monitoring of cloud coverage both during daylight and night time, allowing a continuous monitoring of the cloudiness for statistical and forecast purposes. The all-sky camera is equipped with interfaces for the PC connection and data storage. Images can be saved at different time intervals; from the stored images a log of the cloudiness of the sky in the field of view of the ASTRI SST-2M prototype will be created.

The Sky Quality Meter - LE (SQM) measures the brightness of the night sky in magnitudes per square arcsecond with a 10\% precision ($±0.1$ mag/arcsec$^2$). The SQM is sensitive only to visual light and the model installed at SLN presents a Half Width Half Maximum of the angular sensitivity equal to ~10$^\circ$. The system returns integral information about background light intensity inside the FoV on demand up to a frequency of 1 Hz. The values of the measured sky brightness wil be registered in the log file of the observation performed with the ASTRI SST-2M prototype.

The calibration of a Cherenkov telescope includes \-se\-ve\-ral items and tools. The camera of the ASTRI SST-2M telescope is equipped with internal devices \cite{bib:Catalano} which measure the mirrors alignment, the pointing accuracy and the camera gain monitoring. Nevertheless, the absolute calibration of the overall system can take advantage from auxiliary instrumentation external to the telescope.

 \begin{figure}[t]
  \centering
  \includegraphics[width=0.45\textwidth]{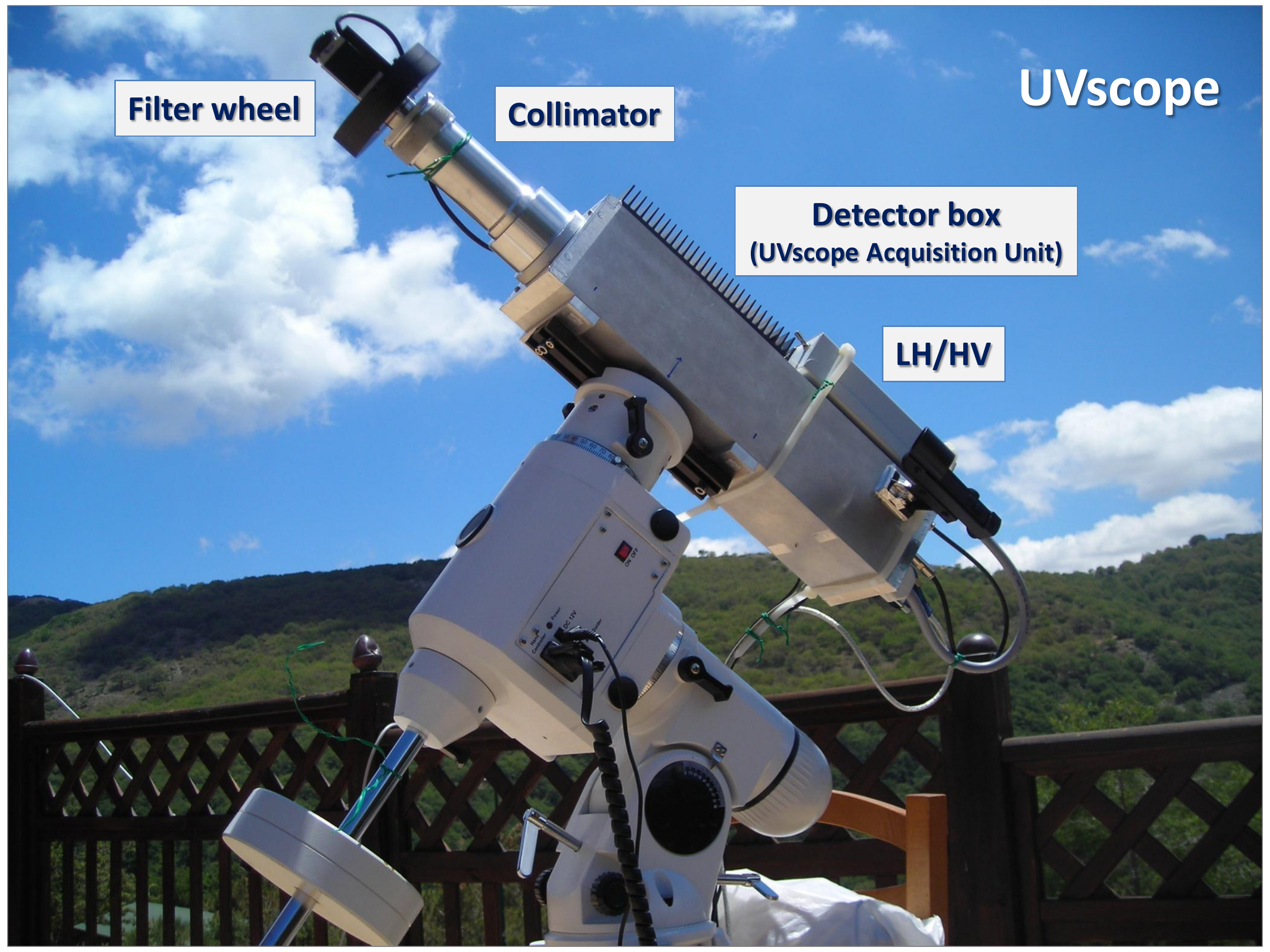}
  \caption{UVscope on its mount during the 2009 campaign in Contrada Pomieri, Italy (1335 m a.s.l.) devoted to the study of NSB and atmospheric transparency at several wavelengths making use of a motorized filter wheel \cite{bib:Maccarone_NIMA}.}
 % For this application, we  equipped the UVscope collimator with a motorized wheel containing narrow-band (337, 355, 391 and 420 nm) and wide-band (M-UG6) filters.
  \label{figure.4}
 \end{figure}

 \begin{figure}[t]
  \centering
  \includegraphics[width=0.4\textwidth]{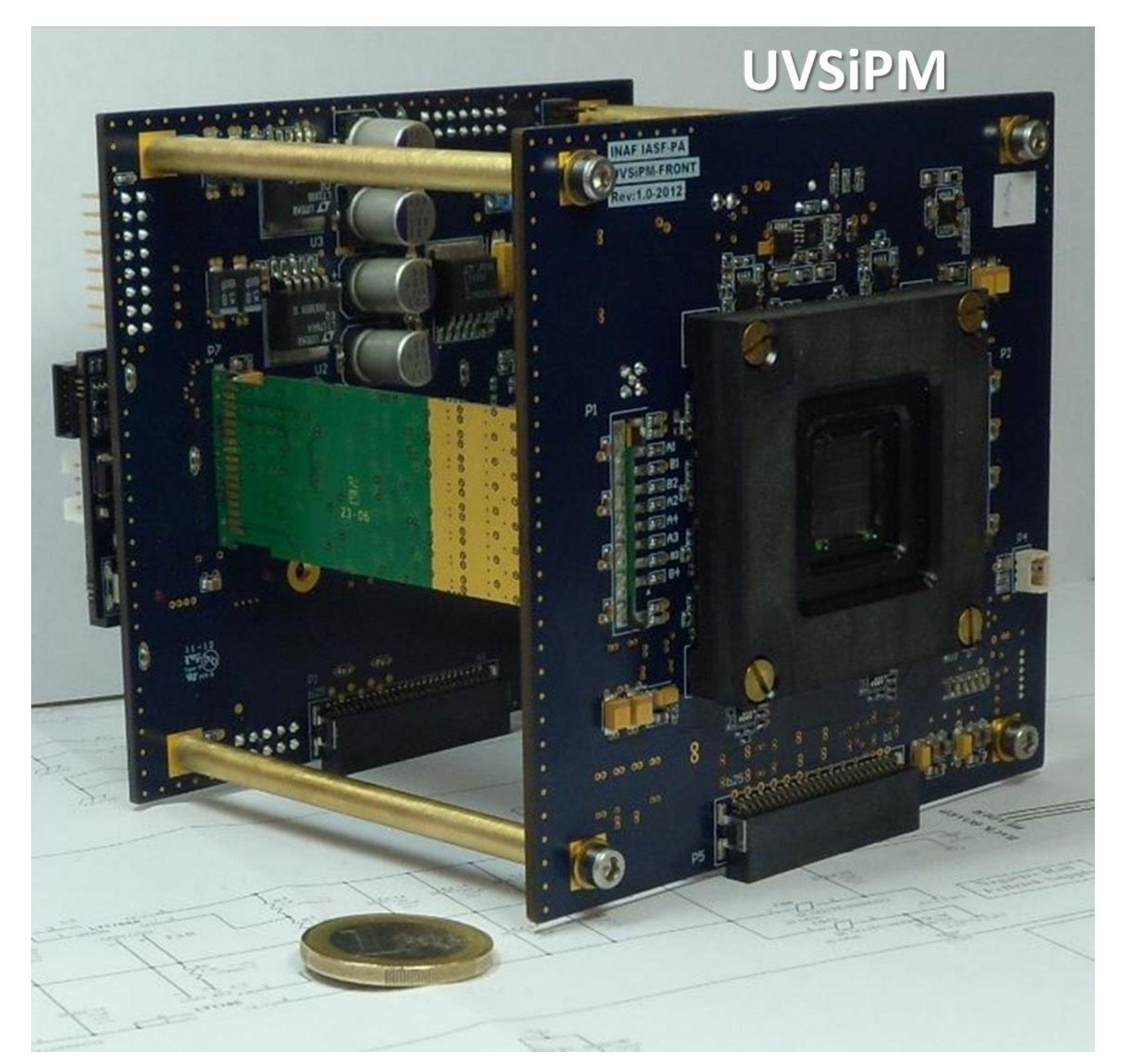}
  \caption{Inside UVSiPM: front view of the acquisition unit with the
socket for the SiPM sensor \cite{bib:Sottile_SCINEGHE}.}
  \label{figure.5}
 \end{figure}

In the case of ASTRI SST-2M installed at SLN, the absolute calibration of the whole telescope system will be mainly performed with the support of the complex UVscope \cite{bib:Maccarone_NIMA} and UVSiPM \cite{bib:Sottile_SCINEGHE}, two portable multi-pixels photon detectors operating in single counting mode, entirely developed at the IASF-Palermo/INAF Institute.  The UVscope (Fi\-gu\-re \ref{figure.4}) sensor is a multianode photo-multiplier (wavelengths range 300-650 nm), while the UVSiPM (Figure \ref{figure.5}) sensor is a Silicon photomultiplier (wavelengths range 320-900 nm) of the same model of sensors that will fill the camera at the focal plane of the ASTRI SST-2M prototype. The two detectors forming  UVscope-UVSiPM will be configured with proper entrance pupil and collimator length in order to obtain the same FoV; moreover, both the instruments will be completed with equal calibrated filters inside their collimators. Both  UVscope and UVSiPM, mounted on a motorized SmartStar MiniTower Pro, will be moved contemporarily pointing the same source without any \-in\-ter\-fe\-rence between them nor with the ASTRI SST-2M operations. Several kinds of acquisitions are foreseen; as an example, UVscope-UVSiPM can measure the diffuse emission of the Night Sky Background (NSB) in the ASTRI SST-2M field of view, allowing a real time gain monitoring. During clear nights, both ASTRI SST-2M and UVscope-UVSiPM would simultaneously "track" a reference star pointing at the RA-Dec star position and following it at different elevation angles; thanks to the accurate calibration of UVscope-UVSiPM performed in lab, the yielded flux profiles will allow us to determine the total atmospheric attenuation and the absolute calibration constants for the prototype. Last but not least, UVscope-UVSiPM will be used, simultaneously with ASTRI SST-2M, to observe a ground light source of well-known properties; the comparison of the separated but simultaneous acquisitions will result in the determination of the spectral response of the whole ASTRI SST-2M telescope, including both optics and camera \cite{bib:Maccarone_TN_Calib}.

\section{Conclusions}
ASTRI SST-2M is an end-to-end prototype of the Small Size Telescope compliant with the CTA requirements.

The technological solutions adopted and the expected performance of ASTRI SST-2M will be verified through a careful calibration and scientific data acquisition phase that will be conducted at the Serra La Nave observing station where the prototype will be installed in 2014.

Serra La Nave represents the best INAF Italian site for Cherenkov astronomy and, after the installation of ASTRI SST-2M, it will host the biggest optics/UV telescope in the National territory.\\

\vspace*{0.5cm}
\footnotesize{{\bf Acknowledgment:}{This work was partially supported by the ASTRI "Flagship Project" financed by the Italian Ministry of Education, University, and Research (MIUR) and led by INAF, the Italian National Institute of Astrophysics. We also acknowledge partial support by the MIUR 'Bando PRIN 2009'.}}

\end{document}